\documentclass[preprint,aps,draft]{revtex4}

\usepackage{graphicx}
\usepackage{dcolumn}
\usepackage{bm}

\begin{document}

\title[Some Remarks about Variable Mass Systems]{Some Remarks about
Variable Mass Systems}

\author{Mario J. Pinheiro}
\email{mpinheiro@ist.utl.pt}

\address{Department of Physics and Centro de F\'{i}sica de Plasmas,
Instituto Superior T\'{e}cnico, Av. Rovisco Pais, \& 1049-001
Lisboa, Portugal}

\date{\today}%





\begin{abstract}
We comment about the general argument given to obtain the rocket
equation as it is exposed in standard textbooks. In our opinion,
it can induce students to a wrong answer when solving variable
mass problems.
\end{abstract}

\maketitle

It is frequently found in standard textbooks
~\cite{Sommerfeld,Serway,Allonso,Nussenzveig} a not enough careful
demonstration of the rocket equation. The argument usually
presented is misleading (although the final result is fortunately
correct) and the consequences of the application of this argument
to other situations, like the problem of a tank firing a
projectile, can induce a wrong answer. One exception to this
general presentation is found in ~\cite{French}, which carefully
call attention to the approximate character of the demonstration.

The correct argument to deduce the rocket equation (established by
Ivan Vsevo\-lodovich Meshchersky in 1897) should be: consider a
rocket with mass m and velocity v relative to an inertial frame at
instant of time t (usually is the Earth). At time $t+\Delta t$, a
quantity of matter of mass $\delta \mu$ (burnt fuel) have been
ejected with velocity $-\mathbf{v}_e$ relative to the rocket (the
velocity of the ejected fuel is usually assumed constant), while
the remaining mass of the rocket ($m-\delta \mu$) has its velocity
increased by $\mathbf{v}+\delta \mathbf{v}$. But, at instant of
time $t + \Delta t$ (and in contradistinction to the general
reasoning) the mass $\delta \mu$ has then a velocity
$\mathbf{v}+\delta \mathbf{v}-\mathbf{v}_e$ relatively to the
ground$\!$. That is, the momentum before and after firing is
\begin{equation}\label{}
\begin{array}{c}
  \mathbf{p}(t)=m\mathbf{v}\\
  \mathbf{p}(t+\Delta t)=(m-\delta \mu)(\mathbf{v}+\Delta
  \mathbf{v})+ \delta \mu (\mathbf{v} + \Delta \mathbf{v} -
  \mathbf{v}_e).
\end{array}
\end{equation}
The change of the linear momentum in the interval of time $\Delta
t$ is due to the action of an external force $\mathbf{F}^{ext}$ -
supposed to be only the gravitational force. When obtaining
$\Delta \mathbf{p}$ all the terms are cancelled out - and there is
no need to justify the neglect of the higher order term $\delta
\mu \Delta \mathbf{v}$. The fundamental equation of dynamics gives
\begin{equation}\label{}
\Delta \mathbf{p}=\mathbf{p}(t + \Delta t)-\mathbf{p}(t)=
\int_0^{\Delta t} \mathbf{F}^{ext} d t = -m \mathbf{g} \Delta t.
\end{equation}
In the limit $\Delta t \longrightarrow 0$ and projecting the
vectorial equation in a vertical axis oriented along $\mathbf{v}$,
we obtain
\begin{equation}\label{rocket1}
-m g = -v_e \frac{d \mu}{d t} + m \frac{d v}{d t}.
\end{equation}
As the total mass (rocket + combustible) is constant, $M = m +
\mu$, then $d M = 0 = d m + d \mu$ and the known form of the
equation is retrieved
\begin{equation}\label{rock2}
-m g = m \frac{d v}{d t} + v_e \frac{d m}{d t}.
\end{equation}
The pedagogical error introduced in the general argument displayed
in standard textbooks will induce a serious error when students
are reasoning about variable mass systems, like the tank firing a
projectile. To illustrate this point better, let's consider a
canon installed on a tank moving without friction over an
horizontal track. The canon and tank with total mass M are both
moving initially with velocity u when an projectile with mass m
(not included in M) is fired with velocity $v$ relative to the
tank. Consider the axis of the canon and the track axis are both
on the same vertical plan and the canon do an angle $\alpha$ with
the horizontal plan, with $\alpha$ being an acute angle relatively
to the direction of displacement of the tank. Considering that the
net force along the horizontal axis is zero, there is conservation
of momentum along this direction.

There is a general believe that there is an ambiguity in the
statement of the problem, because some people argue that
$\mathbf{v}$ should be relative to the velocity of the tank before
ejection and other people argue that $\mathbf{v}$ should be the
velocity v relative to the tank after ejection. Of course, there
is no ambiguity at all: the velocity v of the projectile relative
to the tank before ejection is null $!$. When stating the
conservation of momentum, there is only one possible equation:
\begin{equation}\label{}
\begin{array}{c}
  \mathbf{p}(t)=(M+m) u\\
  \mathbf{p}(t+\Delta t)=M(u + \delta u)+ m (v \cos \alpha+ u + \delta u).
\end{array}
\end{equation}
The recoil speed {\it should be}
\begin{equation}\label{rec}
\delta v = -\frac{m v \cos \alpha}{M+m}.
\end{equation}
and not, $\delta u=-\frac{m v \cos \alpha}{M}$. Consequently, the
range of the projectile in the Earth frame should be
\begin{equation}\label{}
x_A = \frac{2v \sin \alpha}{g} \left(u+ \frac{M v \cos
\alpha}{M+m} \right).
\end{equation}
and no other solution is correct. The problem is clear. Of course,
if $m\ll M$ the error is negligible, but otherwise it is not. This
kind of generally accepted reasoning should be corrected because
it could lead students (and possibly military engineers...) to the
wrong answer.


\end{document}